# Robustness of FCS (Fluorescence Correlation Spectroscopy) with Quenchers Present

*Hima Nagamanasa Kandula,[1] Ah-Young Jee,[1] Steve Granick,[1,2] ***

[1]Center for Soft and Living Matter, Institute for Basic Science (IBS), Ulsan 44919, South Korea

[2]Department of Chemistry, Ulsan National Institute of Science and Technology (UNIST), Ulsan 44919, South Korea

Fluorescence correlation spectroscopy, Fluorescence Quenching, Diffusion coefficient.

## Abstract

Inspired by recent publications doubtful of the FCS technique, we scrutinize how irreversible ("static") and reversible ("dynamic") quenching can influence the interpretation of such data. Textbook presentations often emphasize only how to analyze data in extremes, absence of quenching or the presence of substantial quenching. Here we consider intermediate cases where the assessment of photophysics (static quenching, blinking-like triplet state relaxation) influence on autocorrelation curves can be delicate if dye-labeled objects diffuse on comparably-rapid time scales. We used the amino acid, tryptophan, as the quencher. As our example of small-molecule dye that diffuses rapidly, we mix quencher with the fluorescence dye, Alexa 488. The translational diffusion coefficient, inferred from fit to the standard one-component Fickian diffusion model, speeds up without loss of quality of fit, but quenching is reflected in the fact that the data become exceptionally noisy. This reflects the bidisperse population of quenched and unquenched dye when the time scales overlap between the processes of translational diffusion, quenching, and blinking. As our example of large-molecule dye-labeled object that diffuses relatively slowly, we mixed quencher with dye-labeled BSA, bovine serum albumin. Diffusion, static quenching and blinking time scales are now separated. In spite of quenching contribution to the autocorrelation function when the delay time is relatively short, the inferred translational diffusion coefficient now depends weakly on presence of quencher. We conclude that when the diffusing molecule is substantially slower to diffuse than the time scale of photophysical processes of the fluorescent dye to which it is attached, influence of quenching is self-evident and the FCS autocorrelation curves give appropriate diffusion coefficient if correct fitting functions are chosen in the analysis.

## I. Introduction

In the vast field of single-molecule fluorescence imaging[1-2], FCS (fluorescence correlation spectroscopy) is used frequently as a sensitive reporter of dynamical processes, especially translational diffusion,[3-6] because it presents large advantages over competing approaches. Not only can it give access to processes as fast as ~100 ns, which is orders of magnitude more rapid than objects can be imaged in real space, but also the analysis of FCS data gives an ensemble-average of many molecules each of them with single-molecule sensitivity, so the experiment is sensitive to nM or less amounts of sample[4]. A concern is that fluorescence intensity fluctuations, which are the raw data for this experiment, can also be influenced by photophysical processes such as photobleaching, quenching, and triplet state dynamics. This was considered in scattered early literature when the FCS technique was developed[7-10] but with emphasis, at that time, on photophysical changes so large that corruption of the data would be obvious to the eye. Here we revisit the problem with the intent of clarifying the robustness of this method to characterize translational diffusion with particular focus on rather small quencher concentrations.

We are inspired by recent remarks critical of the FCS technique[11-14] in the context of enhanced enzyme diffusion during catalysis. The suggestion was made that the observed increase in the diffusion coefficient of enzymes during their catalysis[15-17] could have a significant contribution from photophysical effects, namely fluorescence quenching either by the substrate (or catalyst) or products created in the due course of catalysis rather than entirely from enzyme motion. Fischer and coworkers refer to "effect and artifacts in FCS measurements of protein diffusion". Zhang and Hess refer to "possible sources of artifacts in the widely employed FCS measurements". Ross, Sen and coworkers refer to "artifacts in FCS measurements due to its sensitivity to environmental conditions." These critiques were made without modeling to test the claim, however.

The principle of FCS experiment is to illuminate a roughly femtoliter-sized observation window, produced by confocal, two-photon, or super-resolution optics, while analyzing the emitted fluorescence intensity fluctuations. Among the data that can be inferred are translational diffusion, rotational diffusion, and interaction rates such as from proteins[1,6] and DNA assembly[18-19] that are produced by variations in the dye's local chemical environment.

## II.    Results and Discussion
### i.    The notions of static and dynamic quenching.

There is no ambiguity how to analyze such data when time scales are well separated, for example when the time scale for a molecular dye to diffuse across the observation window ($\approx$10 μs) much exceeds that of the fluorescence lifetime decay ($\approx$ns) and the latter is the only relevant photophysical process (Fig. 1a, case A). Alternatively, fluorescence may be lost by bleaching or encounter with a quencher molecule, so-called "static quenching." In this case, loss of fluorescence is irreversible within the observation window (Fig 1a, case B). The option also exists of intermittent fluorescence loss by blinking which could be due to triplet relaxation or transient encounter with a quencher molecule to produce "dynamic quenching" (Fig 1a, case C). Triplet state relaxation is a common form of blinking observed in fluorophores arising from intersystem crossing from excited singlet state to a triplet state, a non-fluorescent state before a molecule goes back to its ground state. Textbook presentations of the FCS technique often emphasize Case A, but it is the only one in which the intensity of fluorescence tracks directly the presence of dye in the observation window

These processes, Cases A, B and C, can all contribute to G(t), the autocorrelation curve when intensity time traces are correlated temporally.

## ii.   Autocorrelation curves when static and dynamic quenching contribute.

To analyze cases B and C, we took the approach of generating G(t) synthetically under various model scenarios, supposing simple known functional forms for each one. It is known that for Case A[4]

$$G(t)_D = \frac{1}{N}\left(1 + \frac{t}{\tau_D}\right)^{-1}\left(1 + \frac{t}{\tau_D \kappa^2}\right)^{-\frac{1}{2}} \qquad (1)$$

where N is the average number of fluorophores in the observation volume, $\tau_D$ is the translational diffusion time, t is the correlation time, and $\kappa$ is the ratio, width to height, of the confocal volume. In our experimental setup we determined the parameters $\tau_D$=41 μs and $\kappa$=4 from FCS measurements of Alexa 488, a dye known for its high photostability, at ~1nM concentration in PBS buffer at pH=7.2. To investigate Cases B and C experimentally, Alexa 488 was mixed with the amino acid tryptophan (Trp), a known quencher[20-21]. We are not aware of prior FCS study of this system. Exploring the joint occurrence of Cases A and B, we simulated the autocorrelation curve for a range of time constants of photo bleaching ($\tau_B$) and the average fraction (B) of the molecules that photobleach:

$$G(t)_{BD} = G(t)_D * G(t)_B$$
$$= \frac{1}{N}\left(1+\frac{t}{\tau_D}\right)^{-1}\left(1+\frac{t}{\tau_D \kappa^2}\right)^{-\frac{1}{2}}\left(1 + B\left(e^{-t/\tau B} - 1\right)\right) \qquad (2)$$

For B > ~0.7 and $\tau_B$ < ~5 µs, G(t)$_{BD}$ exhibits two distinct decay regimes, as was known from prior study[9, 22-23]. However, for small B and $\tau_B \approx \tau_D$ the data can be fit satisfactorily to Eq. 1 within typical experimental uncertainty except with a faster decorrelation time and accordingly, more rapid implied diffusion. Thus extracted or apparent diffusion time ($\tau_{DA}$) decreases by a factor up to 1.5 when B = 0.3 and $\tau_B$ = 30 µs (Fig. 1b and inset). Physically, this occurs because when a molecule loses fluorescence despite being within the observation window, naïvely Eq. 1 interprets such data as faster diffusion[9, 23].

Considering Case B, we multiply Eq. 1 by the expression often used to describe blinking caused due to triplet relaxation (Case C)[8, 23],

$$G(t)_{TD} = \frac{1}{N}\left(1+\frac{t}{\tau_D}\right)^{-1}\left(1+\frac{t}{\tau_D \kappa^2}\right)^{-\frac{1}{2}}\left(\frac{1-T+T exp^{-\frac{t}{\tau_T}}}{1-T}\right) \qquad (3)$$

where the triplet life time ($\tau_T$) and the fraction of the molecules in triplet state (T) are the variables in this equation. Equivalently, in the context of quenching, they represent the lifetime of the photo-induced complex and the probability to form the complex, respectively. The influence on FCS curves of blinking can be guessed by taking a closer look at the intensity profile (Case C in Fig 1a). Naively one would expect G(t) to have two characteristic times, one corresponding to transient off-time, at which intensity is zero but the fluorophore remains in the observation volume, and the other corresponding to when the fluorophore actually leaves the observation volume by diffusion. However, two decay times become obvious to the eye only for large T and $\tau_T \ll \tau_D$, just as in Case 2. As dynamic complexes are typically short-lived for commonly used stable dyes like the Alexa family, it is reasonable to consider $\tau_T = 4\mu s$ for Alexa 488, in which case the autocorrelation curve appears to only shift progressively to faster decay times as seen in the decrease in $\tau_{DA}$ (Fig. 1b and its inset).

Parametric studies of Cases B and C demonstrate the importance of separation of time scales in order to discriminate between photophysical processes and diffusion, and hence unambiguously extract the relevant diffusion time scales from FCS curves. To check this, we compare $G(t)_{TD}$ for a small-molecule fluorophore ($\tau_T = 4\mu s$ and $\tau_D = 40$ $\mu s$) and when the fluorophore is attached to a macromolecule thereby increasing the diffusion time ($\tau_T = 4\mu s$ and $\tau_D = \sim 240$ $\mu s$). Clearly for macromolecules two decay profiles become evident even at very small T (T=0.2). More importantly, $\tau_{DA}$ shows smaller deviations from $\tau_D$ even at larger values of T (inset of fig 1c). The relevance to experimental systems is that small ($\sim 100\mu M$) concentrations of quencher in quencher-dye systems, i.e. small but finite B and T in Eq. 2 and 3, shift intensity-intensity autocorrelation curves, as supposed in Case A, the case for which there is no quenching at all.

However, to separate photophysical processes in an actual experimental situation is a delicate matter. FCS measurements were performed using a commercial Leica SP8 with continuous laser excitation at $\lambda = 488$ nm, and at low laser powers to avoid photo bleaching, with a 100X oil immersion objective, and detection using two avalanche photodiodes. Cross-correlation removed after-pulsing that would otherwise corrupt measurements below ~μs[4, 24].

### iii.    Experiments using a small-molecule moving object.

With this in mind, we mixed ~nM concentrations of our standard dye (Alexa 488) with quencher (Trp) at concentrations varied from 10 μM to 1 mM in PBS buffer at pH=7.2. These low quencher concentrations, although not explored clearly in earlier literature to the best of our knowledge, are important in the burgeoning field of enzyme-substrate systems.

At concentrations above 100 μM Trp, the fluorescence intensity decreases rapidly reflecting increased quenching (Fig. 2a), and this is accompanied by faster fluorescence decay lifetime, $\tau_F$. It is known that diminished intensity can stem from both static and dynamic quenching while speeded-up fluorescence decay lifetime only reflects dynamic quenching[20]. Therefore, our observation of larger change for $(I_0/I)$ versus concentration as compared to $(\tau_0/\tau)$ versus concentration indicates that static and dynamic quenching coexist even at low concentrations of Trp (Fig. 2b). This is consistent with what is known for Trp-Alexa 488 system at concentrations above 1mM of Trp[20].

The corresponding G(t) for this data (Fig. 3a) are not significantly perturbed in shape. This reflects the close proximity of $\tau_T$ and $\tau_D$ and small concentrations of Trp used. Nevertheless, with increasing Trp concentration, the curves become nosier beyond ~10 μs lag time. This is accompanied by shift of G(t) to more rapid decay times (Fig. 3b). These changes are likely due to increased probability of blinking with increasing quencher concentration. In fact, T and $\tau_T$ can be extracted by fitting $G_{TD}$ (Eq.3) to G(t)s in Fig 3a. As expected, we find that T increases monotonically with Trp concentration and $\tau_T$ displays fluctuations (Fig. 3c and d).

Another striking feature in Fig 3a is decrease of G(0) with increasing Trp concentration. It is known that the value of G(t) at vanishingly small delay times, G(0), is inversely proportional to the average concentration of fluorophore[4]. Therefore, the changes observed here naively suggest a systematic increase in the apparent concentration of the fluorophore with increasing Trp (Fig. 3e), but this is not reasonable physically as the actual concentration of fluorophore should either remain constant or diminish slightly due to static quenching. Autofluorescence of tryptophan does not explain this change since the excitation and emission wavelengths for intrinsic tryptophan fluorescence ($\lambda_{exe}$ ~280 nm and $\lambda_{emi}$ ~350 nm) are well below those used for our experiments ($\lambda_{exe}$ ~488 nm and $\lambda_{emi}$ ~500 nm to 550 nm). One contribution to the observed decrease in G (0) with increasing quencher concentration is likely to be reduced instantaneous fluorescence intensity with increasing quencher concentration without large change in the background level. Furthermore, in the literature, such apparent changes in the concentration are also thought to arise from changes in intensity fluctuations due to changes in refractive index, viscosity or pH[25-27]. While the exact physical process that leads to this apparent increase in concentration in our system remains unknown, our experiments clearly show that even for a simple dye solution, presence of small amounts of quencher can lead to spurious estimates of dye concentration even when the profile of G(t) is visually unchanged.

Moving to diffusion, naïve inspection would conclude, from deducing $\tau_D$ from Eq. 3, that it increases slightly with increasing quencher concentration (Fig. 3f), whereas correlation decay appears to speed up (Fig. 3b), which is not self-consistent. This is because we used too many free parameters, in this case 4 parameters (N, T, $\tau_T$ and $\tau_D$). It does not seem practical to fit FCS data with so many parameters reliably, even if one has insight into the various physical processes that may contribute. The difficulty is that the time scales are all similar (μs): those of triplet, dye-quencher complexes and translational diffusion of dye.

## iv.    Experiments with a larger moving object

In a system better suited to separating these time scales, experiments were performed with macromolecules that diffuse more slowly because of larger size. Bovine serum albumin (BSA) labeled with Alexa 488 was purchased (ThermoFisher Scientific). Samples at concentration ~2 nM were dissolved in PBS buffer (pH 7.2) and filtered using 100 kDa centrifugal filters (ThermoFisher Scientific) to remove possible aggregates.

First, we measure the fluorescence decay time in pure PBS to determine fluorescence properties of Alexa when it is chemically attached to BSA. The decay profile changes from single exponential to two exponentials and the fluorescence lifetime of Alexa decreases by a factor of 1.3 (Fig 4a). Further small changes are observed when Trp is present (inset of Fig 4a). We now move to G(t) of BSA in PBS, shown in Fig 4b. We find that simple $G_{TD}$ (Eq.3) cannot describe this correlation curve, and therefore use a model with two independent quenching events,

$$G(t)_{T2D} = \frac{1}{N}\left(1+\frac{t}{\tau_D}\right)^{-1}\left(1+\frac{t}{\tau_D\kappa^2}\right)^{-\frac{1}{2}}\left[\Sigma_{j=1}^2 T[j]\left(exp^{-\frac{t}{\tau_{T[j]}}}-1\right)+1\right] \qquad (4)$$

Using $G(t)_{T2D}$, we obtain $\tau_{T1}$ = ~4 μs corresponding to triplet blinking and another time constant, ~150 μs, corresponding to internal motion of the protein[28-29].

Therefore, to analyze correlation curves of BSA samples with and without Trp, we use $G_{T2D}$. Just as for the case of free dye, from G(0) we infer increase in apparent concentration of protein with increasing Trp concentration. This is accompanied by increasing noise in G(t), but noise vanishes at time scales slower than ~10 μs and does not contribute to the slower decay time that underlies the inference of BSA translational diffusion of BSA ($\tau_D$ = ~250 μs, Fig 4c). Therefore, for Trp concentration < 1mM the diffusion coefficient remains largely unchanged by quencher. However, it changes drastically at concentrations above 1mM (Fig. 4d). Notably, for Trp = 1mM, unlike in free dye-quencher system, a slower decay is obvious even in the raw FCS curve, consistent with larger $\tau_D$ extracted from fitting (Fig 4c and d). Therefore, though additional time scales influence the correlation curve, these experiments with Alexa-labelled BSA demonstrate that the influence on diffusion of quenching can be analyzed tractably, by separating the time scales of the photo physical process and diffusion in the system under study.

## III.    Conclusion.

In the years since the pioneering invention and development of the FCS technique[3, 30], FCS has frequently come to be used as a turnkey experiment, sometimes without considering how common photophysical process stimulated by the presence of a fluorescence quencher may affect the intensity-intensity autocorrelation curves that constitute the raw data for this experiment. On the one hand, this study shows that the presence of even small amounts of photophysical processes, typically assumed to be too small to significantly perturb correlation curves, can indeed strongly influence inferred quantities such as the translational diffusion coefficient and the average concentration of fluorophores, especially when the time scales overlap (Fig 3). On the other hand, it shows that dynamical properties like diffusion time scales and internal dynamics of a given protein can be extracted with reasonable experimental accuracy, uncertainties of at most ~15 percent for a range of concentrations under the conditions modeled here, if the time scales of various processes are well separated (Fig 4).

A useful observation is that in the presence of quencher, dye concentration is no longer faithfully determined from the inverse of G(0); therefore, unanticipated changes of G(0) could possibly be used as an indicator for the presence of quenching. Further, using functional forms with the appropriate microscopic details can provide insight into novel dynamical features in system like proteins, thereby making this technique powerfully-useful as proposed long ago[7-10], provided that correct fitting functions are chosen to analyze the autocorrelation curves.

Looking beyond the scope of this study, it will be interesting in the future to likewise consider the effect on autocorrelation curves of fluorescence enhancers.

ACKNOWLEDGMENT

This work was supported by taxpayers through the Korean Institute for Basic Science, project code IBS-R020-D1.

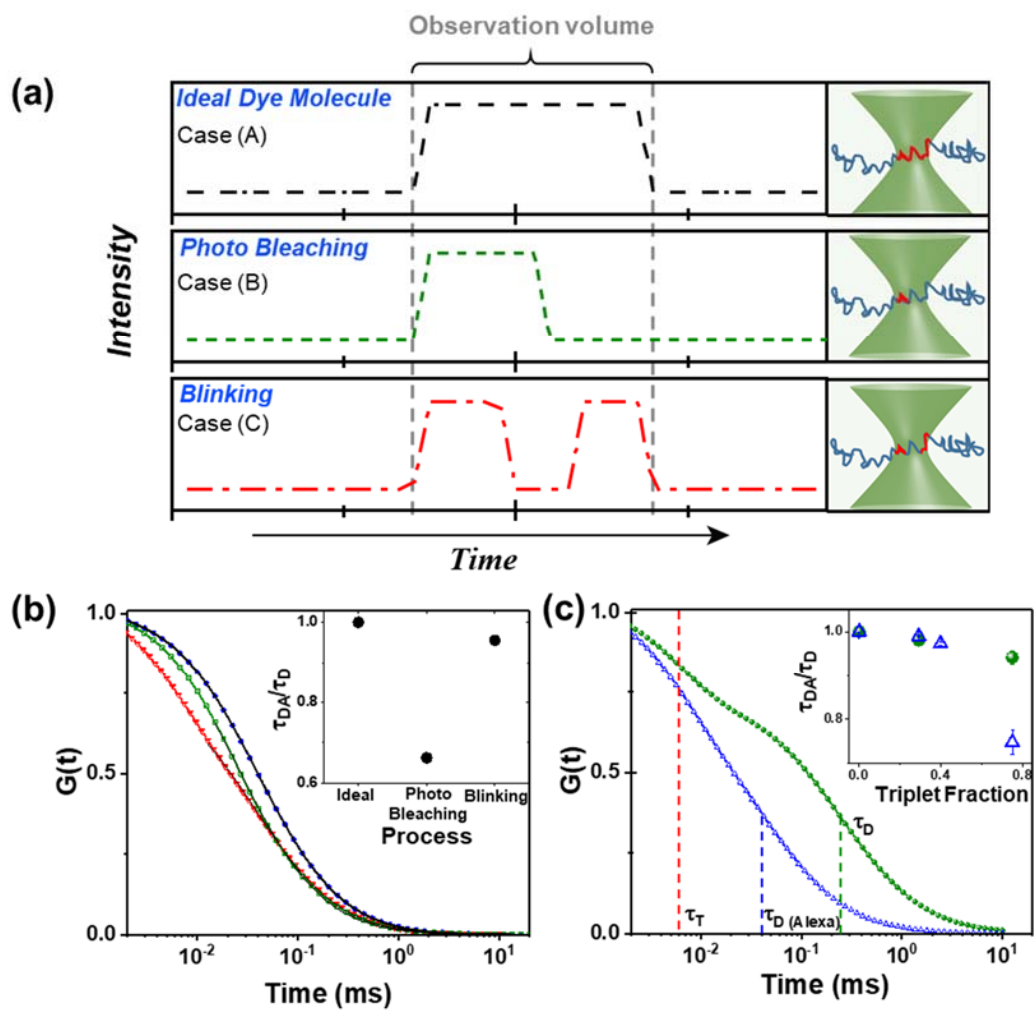

**Figure 1.** Concepts of static and dynamic quenching. **a)** Schematic illustration of fluorescence intensity profiles as a dye molecule passes through the FCS observation volume. The three panels demonstrate intensity profiles for the ideal FCS case (*top panel*), static quenching/photo bleaching (*middle panel*) and blinking that reflects triplet relaxation or dynamic quenching (*bottom panel*). **b)** Computed FCS intensity-intensity autocorrelation curves for a small-molecule system using Eqs. 1, 2 and 3 for cases A, B and C, respectively. Case A: ideal FCS diffusion with $\tau_D$= 41µs (circles). Case B: photo bleaching fraction B=0.3 and time scale $\tau_B$= 30µs. Case C: triplet fraction T=0.3 and time scale $\tau_T$= 4µs (squares and triangles, respectively). The inset shows the apparent

diffusion time ($\tau_{DA}$) extracted by fitting Eq.1 to the curves in B. **c)** The G(t) obtained using Eq. 3 with T=0.293 and $\tau_T$=6 µs but with $\tau_D$s varied, and $\tau_D$= 41 µs and $\tau_D$= 246 µs represent dye solution (blue triangles) and dye-labeled BSA solution (green spheres). The vertical dotted lines call attention to different characteristic time scales. The inset shows the apparent diffusion time as a function of T for dye (triangles) and dye labelled BSA (circles) demonstrating that dynamic quenching has lesser effect due to the separation of time scales.

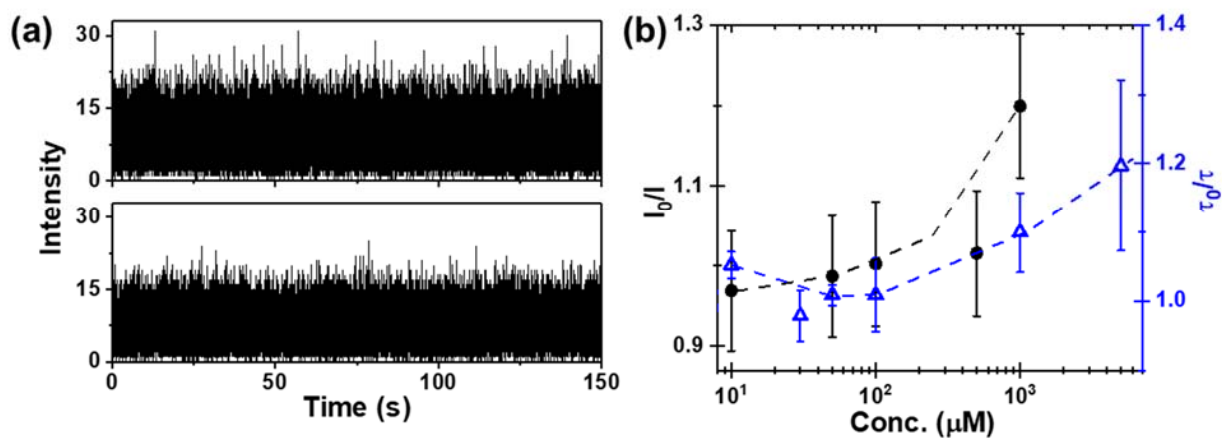

**Figure. 2.** Illustrations how quenching affects fluorescence intensity. The data refer to 1 nM Alexa 488, in PBS buffer at pH = 7.2. Tryptophan is the quencher. **a)** Intensity time traces in the absence of quencher (top panel) and in presence of 1mM quencher (bottom panel). **b)** Fluorescence intensity I relative to the quencher-free state $I_o$ **,** and time scale $\tau$ relative to the quencher-free time scale $\tau_o$, are plotted against quencher concentration. Here, $\tau$ is fluorescence life time obtained from exponential fits to the fluorescence decay with time.

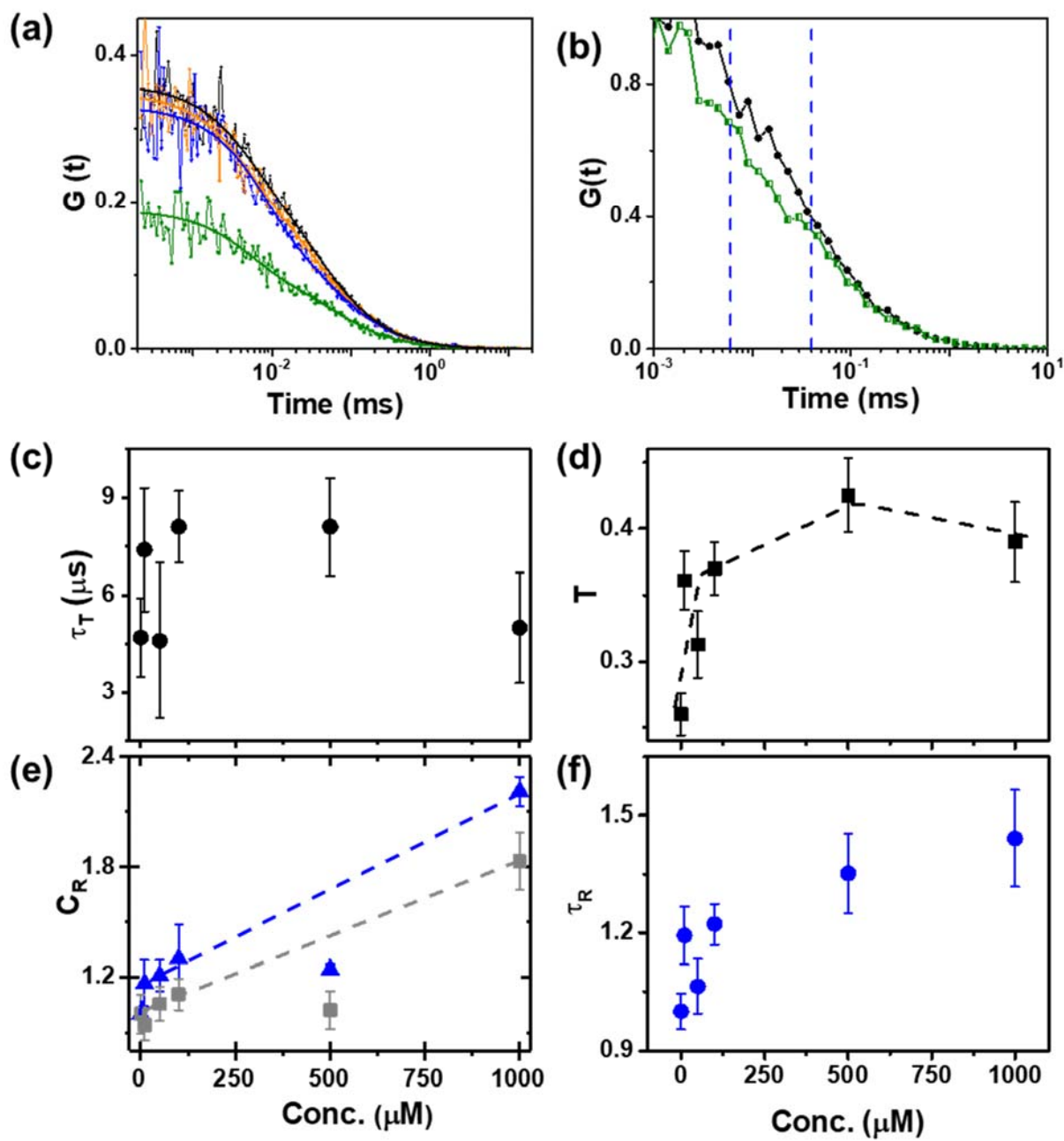

**Figure 3.** Experimental consequences of quenching a small-molecule dye in aqueous solution. The data refer to 1 nM Alexa 488, in PBS buffer at pH = 7.2. Tryptophan (Trp) is the quencher. **a)** Experimental intensity-intensity autocorrelation curves G(t) at various Trp concentrations. Black circles, orange diamonds, blue triangle and green squares correspond to 0, 50 μM, 500 μM and 1mM concentration of tryptophan, respectively. The solid lines correspond to $G_{TD}$ (Eq. 3) fits to G(t). **b)** G(t) normalized by G(t) averaged between 1 and 2 μs, highlighting the fast decay when Trp concentration is high. Black circles and green squares correspond to 0 and 1mM concentration of tryptophan, respectively. As guides to the eye, the vertical dotted lines are drawn at times corresponding to $\tau_T$ and $\tau_D$ in the absence of quencher.. **c)** The triplet time ($\tau_T$), obtained from $G_{TD}$ fits, plotted against Trp concentration. **d)** The triplet fraction (T), obtained from $G_{TD}$ fits, plotted against Trp concentration. **e)** $C_R$, inferred fluorophore concentration in the presence of quencher normalized by inferred fluorophore concentration in the absence of the quencher, obtained in two ways, plotted against Trp concentration. Circles are from $G_{TD}$ fits. Triangles are obtained by averaging G(t) shown in B between 1 and 2 μs. The dotted curves in E are guides to the eye. **f)** $\tau_R$, the ratio $\tau_D / \tau_{D(Trp. =0)}$ is plotted against Trp concentration, demonstrating apparent slowing down.

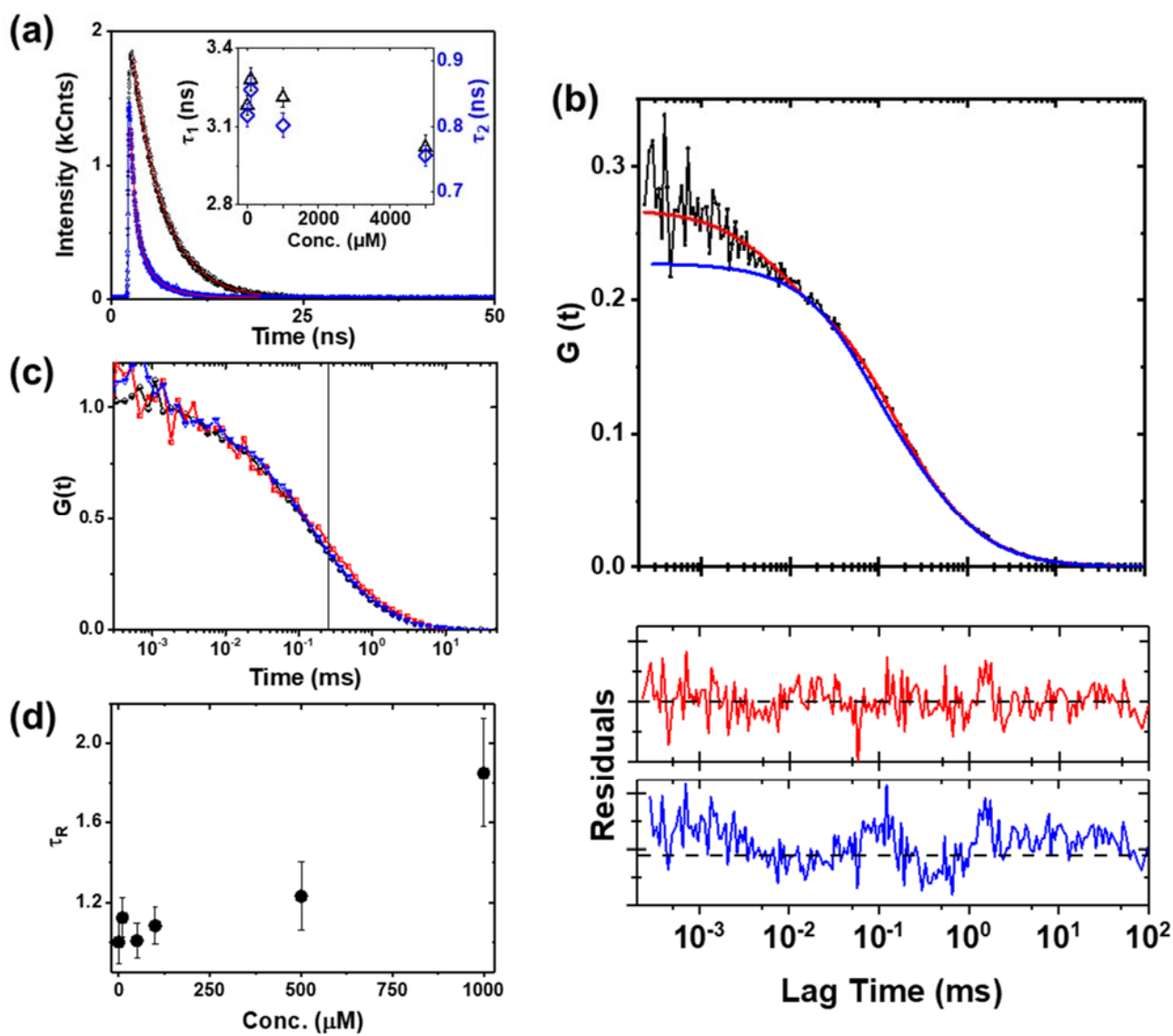

**Figure. 4.** Experimental study of quenching a fluorophore when it is attached chemically to a protein. Trypotophan (Trp) is the quencher. **a)** Intensity plotted against time to determine the fluorescence lifetime of free Alexa 488 (circles) and Alexa 488 attached to bovine serum albumin (BSA). The solid red lines are single and double exponential fits for free Alexa 488 and Alexa 488 attached to BSA, respectively. Inset shows fluorescent lifetimes ($\tau_1$ and $\tau_2$) obtained from double exponential fits to fluorescence decay profiles of Alexa attached to BSA in solutions with different Trp concentrations. **b)** Intensity-intensity autocorrelation function, G(t), of BSA labelled with Alexa 488 in PBS (pH = 7.2). Red and blue solid curves are $G_{T2D}$ and $G_{TD}$ fits, respectively. The bottom panels show residuals to $G_{T2D}$ and $G_{TD}$ fits, in red and blue respectively. **c)** For three Trp concentrations, 0 (black circles), 100 μM (blue triangles) and 1 mM (red squares), normalized G(t) is plotted against time lag; G(t) has been normalized by its average value between 1 and 2 μs, typical smallest time scales observed in G(t) measurements. The curves with black circles, blue triangles and red squares correspond to system with 0, 100 μM and 1mM concentration of tryptophan, respectively. The black vertical line is a guide to the eye, drawn at $\tau_D$ for BSA. **d)** $\tau_R$, the ratio $\tau_D/\tau_{D(Trp. =0)}$ is plotted versus Trp concentration for BSA labelled with Alexa 488, showing no effect exceeding the noise level in the regime of low Trp concentration (up to 0.1 mM), ~15% change just before the point (0.5 mM Trp) of visibly-corrupted G(t), and ~50% change at the point (1 mM Trp) at which G(t) corruption is obvious to the eye..

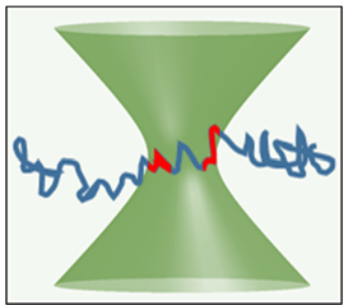